\documentclass[article, twocolumn,
   aps, pra,
  amsmath,amssymb,
  longbibliography,
  ]{revtex4-1}
\usepackage{graphicx,color}
\usepackage{amsmath}
\usepackage{natbib}
\usepackage{epsfig}
\begin{document}

\title{Speed of sound in dense simple liquids}

\author{S. A. Khrapak\email{Sergey.Khrapak@gmx.de} 
}
\affiliation{Joint Institute for High Temperatures, Russian Academy of Sciences, 125412 Moscow, Russia}

\begin{abstract}
The speed of sound of simple dense fluids is shown to exhibit a pronounced freezing temperature scaling of the form $c_{\rm s}/v_{\rm T}\simeq \sqrt{\gamma} +\alpha (T_{\rm fr}/T)^{\beta}$, where $c_s$ is the speed of sound, $v_{\rm T}$ is the characteristic thermal velocity, $\gamma$ is the ideal gas heat capacity ratio, $T$ is the temperature, $T_{\rm fr}$ is the freezing temperature, and $\alpha$ and $\beta$ are dimensionless parameters. For the Lennard-Jones fluid we get $\gamma=5/3$, $\alpha\simeq 7$ with a weak temperature dependence, and $\beta = 1/3$. Similar scaling works in several real liquids, such as argon, krypton, xenon, nitrogen, and methane. In this case, $\alpha$ and $\beta$ are substance-dependent fitting parameters. A comparison between the prediction of this freezing temperature scaling and a recent experimental measurement of the speed of sound in methane under conditions of planetary interiors is presented and discussed. The results provide a simple practical tool to estimate the speed of sound in regimes where no experimental data are yet available. 
\end{abstract}

\date{\today}

\maketitle

\section{Introduction}

The speed of sound is an important property of a fluid, having relation to the equation of state, the bulk modulus, as well as some transport properties such as, for instance, the thermal conductivity coefficient~\cite{Bridgman1923,KhrapakJMolLiq2023}. Similarly to the thermodynamic and transport properties of fluids, there is no general theory and it is unlikely that it will be developed due to the absence of a small parameter~\cite{TrachenkoBook,BrazhkinUFN2012,KhrapakPhysRep2024,PelusoThermo2024}. For this reason, approximate relationships and scalings are particularly important in understanding the properties of fluids. For some relevant examples, we refer to excess entropy scaling~\cite{RosenfeldPRA1977,RosenfeldJPCM1999,DyreJCP2018,BellJPCB2019}, freezing-temperature scaling~\cite{RosenfeldJPCM2001,Kaptay2005,CostigliolaJCP2018,KhrapakAIPAdv2018}, and freezing-density scaling of transport coefficients~\cite{KhrapakPRE04_2021,KhrapakJPCL2022,KhrapakJCP2022_1,KhrapakJCP2024,HeyesJCP2024,KhrapakJMolLiq2025}. Rosenfeld-Tarazona scaling of thermodynamic properties, its implications and modifications are discussed in Refs.~\cite{RosenfeldMolPhys1998,RosenfeldPRE2000,IngebrigtsenJCP2013,KhrapakPRE02_2015, KhrapakJCP2015,MausbachPRE2018,KhrapakPRE09_2024,KhrapakPOF11_2024}.  

Predicting the speed of sound is particularly challenging. In fluids with very soft repulsive interactions, particularly relevant in the plasma-related context, dispersion relations can deviate from the conventional acoustic ones and hence the sound velocity can diverge~\cite{BausPR1980,OhtaPRL2000,KhrapakPoP2016,KhrapakJETPLEtt2024}. For conventional substances with short-range repulsion and long-range attraction, the speed of sound varies greatly in the vicinity of the critical point. The density dependence of the sound velocity along near-critical isotherms is drastically different from that along highly super-critical isotherms. Even using computer simulations, the prediction of the macroscopic adiabatic speed of sound remains a highly non-trivial task~\cite{BrykSciRep2023}.    

The purpose of this paper is to present a simple empirical approach for estimating the speed of sound in simple dense fluids. First, we consider the Lennard-Jones (LJ) fluid and demonstrate that a freezing-temperature scaling applies to the speed of sound for densities above that at the triple point. Then, this scaling is verified using several liquefied noble gases and molecular liquids. We use this scaling to estimate the sound velocity of methane under extreme conditions and to compare this with the result of a recent experiment.  

\section{Motivation}

A physically motivated model of the adiabatic speed of sound would likely start from the conventional definition~\cite{LL_Fluids}
\begin{equation}\label{definition}
c_{\rm s} = \sqrt{\left(\frac{\partial P}{\partial \rho_m}\right)_S}=\sqrt{\frac{\gamma}{m}\left(\frac{\partial P}{\partial \rho}\right)_T},
\end{equation}
where $P$ is the pressure, $\rho_m=m\rho$ is the mass density, $m$ is the atomic mass, $\gamma=c_{\rm p}/c_{\rm v}$ is the heat capacity ratio, $S$ is the entropy and $T$ is the temperature. We can further divide the pressure into the ideal gas and excess contributions:
\begin{equation}
P = P_{\rm id}+P_{\rm ex}=\rho T\left(1+p_{\rm ex}\right),
\end{equation}
where $p_{\rm ex}$ is dimensionless excess compressibility factor. Then we obtain
\begin{equation}\label{definition1}
c_{\rm s} = v_{\rm T}\sqrt{\gamma\left(1+p_{\rm ex}+\rho\frac{\partial p_{\rm ex}}{\partial \rho}\right)},
\end{equation}
where $v_{\rm T}=\sqrt{T/m}$ is the thermal velocity and $T$ is expressed in energy units ($=k_{\rm B}T$). The excess compressibility factor can be expressed using the interatomic interaction potential $\phi(r)$ and the radial distribution function (RDF) $g(r)$ using the pressure equation~\cite{HansenBook}
\begin{equation}
p_{\rm ex} = \frac{2\pi \rho}{3T}\int_0^{\infty}r^3\phi'(r)g(r)dr.
\end{equation}
Looking for general trends in the behavior of the speed of sound, we are stuck at this point for several reasons. First, the excess compressibility factor $p_{\rm ex}$ is strongly system-dependent. It is a very large positive quantity for soft repulsive potentials operating in plasma-related contexts, such as the screening Coulomb potential, because large distances provide a considerable contribution to pressure~\cite{KhrapakPRE02_2015,KhrapakJCP2015}. It drops for steep repulsive interactions, because in this case nearest-neighbor interactions dominate. A relevant limiting case is the hard-sphere fluid~\cite{Pieprzyk2019}. For conventional systems with long-range attraction, the excess compressibility factor is close to zero in the vicinity of the gas-liquid-solid triple point~\cite{RosenfeldCPL1976}. Second, the dependence of $p_{\rm ex}$ on $\rho$ is clearly non-universal but strongly correlates with the shape of the interatomic interaction potential $\phi(r)$, which can vary greatly from one fluid to another. In addition, the derivative of the RDF with respect to $\rho$ is involved, whose evaluation is highly nontrivial.    

Given all this, another approach is chosen. It is assumed that the speed of sound of dense simple fluids exhibits a freezing-temperature scaling. Using the LJ system as a representative example, its form is chosen to combine simplicity and satisfy the two limiting regimes. The scaling is required to reproduce the ideal gas result in the high-temperature limit and to comply with a quasi-universality of the reduced sound speed at the freezing phase transition.  The emerging scaling based on the analysis of the dense LJ fluid is then tested on several atomic and molecular liquids to demonstrate the extent of its generality.

\section{Speed of sound in the Lenard-Jones fluid}  

The pairwise LJ interaction potential is 
\begin{equation}
\phi(r)= 4\epsilon\left[\left(\frac{\sigma}{r}\right)^{12}-\left(\frac{\sigma}{r}\right)^6\right], 
\end{equation}
where $\epsilon$ and $\sigma$ are the energy and length scales. The conventional LJ reduced units for temperature and density are $T^*=T/\epsilon$ and $\rho^*=\rho\sigma^3$, respectively.

\begin{figure}
\includegraphics[width=8cm]{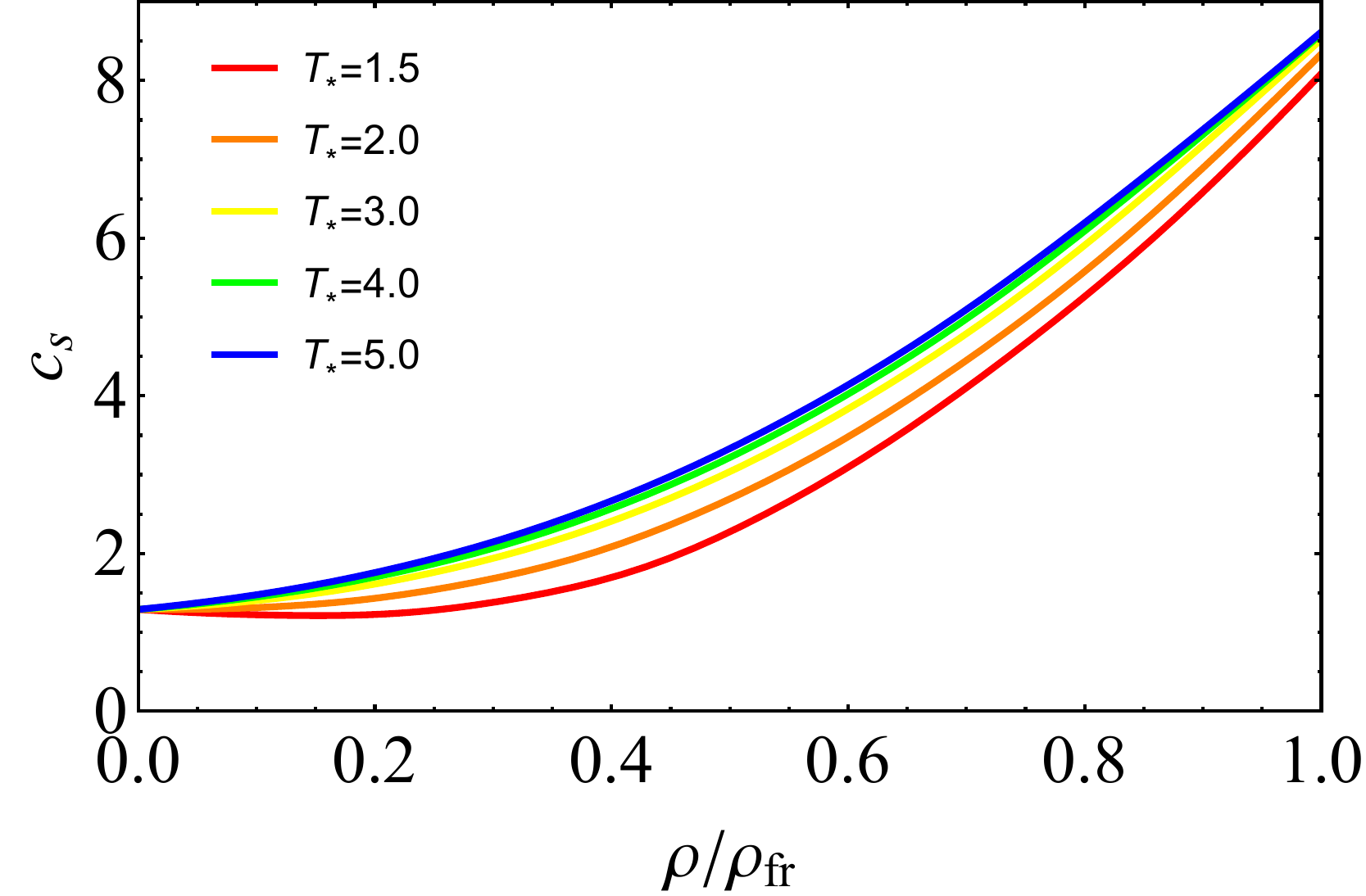}
\caption{(Color online) The speed of sound in units of the thermal velocity of the LJ fluid versus the reduced density $\rho/\rho_{\rm fr}$. The solid curves of different color correspond to different isotherms, see the legend.} 
\label{Fig1a}
\end{figure}

The speed of sound of the LJ fluid has been calculated using the equation of state (EoS) developed by Thol {\it et al.}~\cite{Thol2016}. This EoS provides relatively good accuracy and is convenient in practical implementation. The location of the freezing transition ($\rho^*_{\rm fr}$ and $T^*_{\rm fr}$) is based on the data tabulated in Ref.~\cite{SousaJCP2012}. In view of the success of the freezing-density scaling of transport coefficients~\cite{KhrapakPRE04_2021,KhrapakJPCL2022,KhrapakJCP2022_1,KhrapakJCP2024,HeyesJCP2024}, it is tempting to apply this scaling to the speed of sound. The results of the calculation along five supercritical isotherms, $T^*= 1.5$, 2, 3, 4, and 5 are shown in Fig.~\ref{Fig1a} by solid lines of different color. We do not observe any convincing quasi-universality. The reduced sound speed increases systematically with temperature, although saturation probably occurs as the temperature increases. For this reason, we do not consider freezing-density scaling in the following. Instead, we focus on the freezing-temperature scaling in the high-density regime with densities above the triple point density $\rho^*> \rho^*_{\rm tp}\simeq 0.85$~\cite{SousaJCP2012}.

\begin{figure}
\includegraphics[width=8cm]{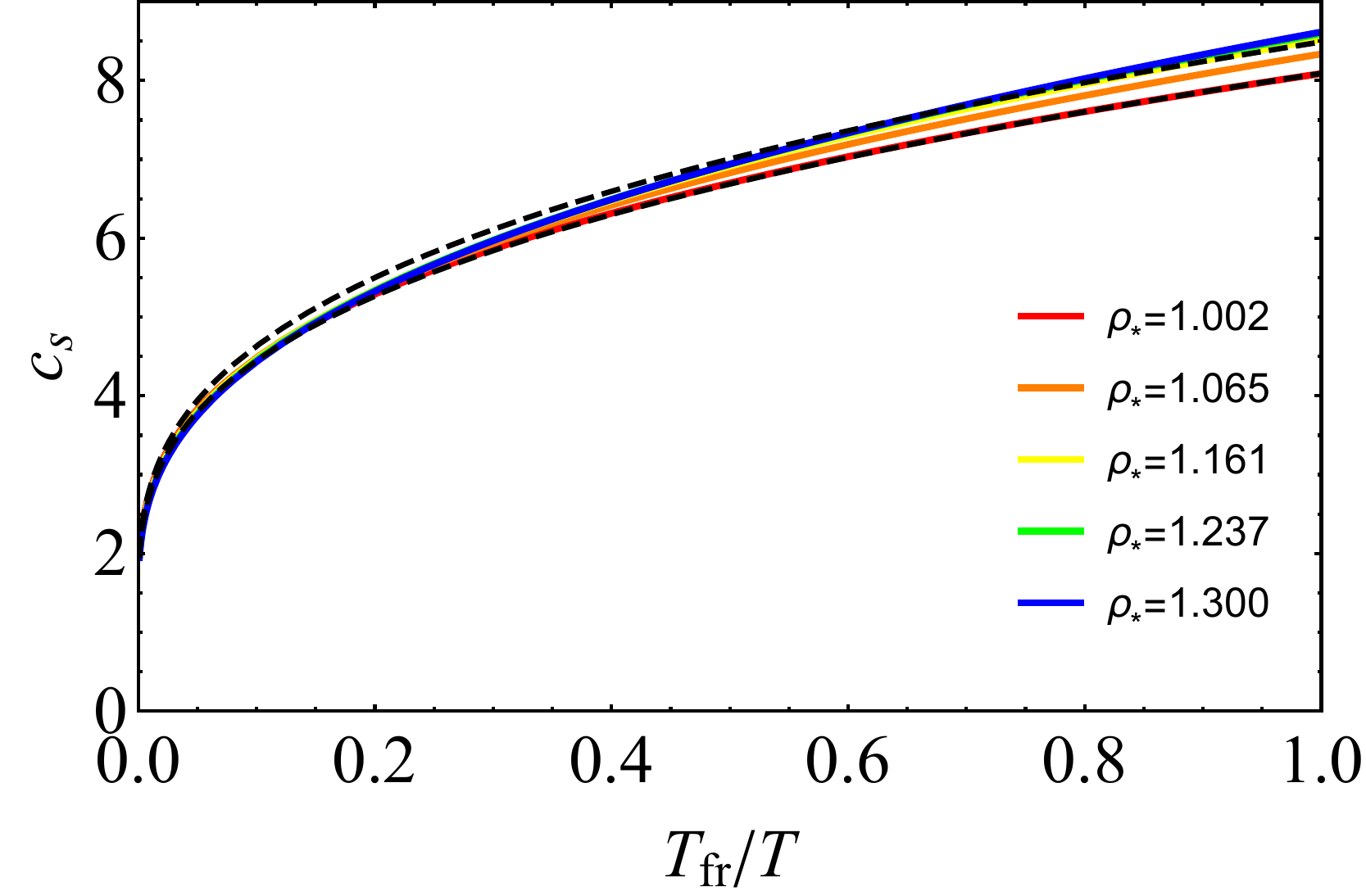}
\caption{(Color online) The speed of sound in units of the thermal velocity of the LJ fluid versus the reduced temperature $T_{\rm fr}/T$. The solid curves of different color correspond to different isochors, see the legend. The two dashed curves correspond to Eq.~(\ref{model}) with $\gamma=5/3$, $\beta=1/3$ and $\alpha = 6.8$ (lower curve) and $\alpha=7.2$ (upper curve).}
\label{Fig1}
\end{figure}

The results of the calculation along five high-density isochores are shown in Fig.~\ref{Fig1}. The densities chosen correspond to the freezing densities at temperatures $T^*= 1.5$, 2, 3, 4, and 5, respectively. In this case, we do observe a convincing quasi-universality, in contrast to the freezing-density scaling. Only when approaching  the freezing transition, the reduced sound velocity demonstrates a slight but systematic increase with temperature. This is a rather general property of various fluids at the freezing line, including the LJ model; see Figs. 3 and 4 from Ref.~\cite{KhrapakPoF2023} as illustrations. For a detailed analysis of the longitudinal and transverse sound velocities at the fluid-solid phase transition in the LJ system, see also Ref.~\cite{KhrapakMolecules2020}. 

Any realistic model of the sound velocity of the dense LJ fluid should satisfy the following two properties. In the high-temperature limit, the ideal gas behavior should be recovered. For the ideal gas, we have $p_{\rm ex}=0$ and thus from Eq.~(\ref{definition1}) it immediately follows that
\begin{equation}
c_{\rm s}=\sqrt{\gamma}v_{\rm T}.
\end{equation}
For the monatomic LJ gas we have $\gamma=5/3$. 

For a given density above $\rho^*_{\rm tp}$, the lowest temperature accessible to the liquid or supercritical fluid is the one at the freezing point. The speed of sound is known to increase very weakly with temperature along the freezing line, with a characteristic value $c_{\rm s}/v_{\rm T}\sim 8$~\cite{KhrapakPoF2023}.

Perhaps a simplest expression consistent with these two properties is just a linear superposition of the ideal gas term and the freezing-temperature scaling of the form
\begin{equation}\label{model}
\frac{c_{\rm s}}{v_{\rm T}}\simeq \sqrt{\gamma} +\alpha (T_{\rm fr}/T)^{\beta}.    
\end{equation}
In general, $\alpha$ and $\beta$ can be treated as adjustable parameters. We shall see in a minute that for the LJ fluid, $\alpha$ increases slightly with the density to allow for a weak temperature dependence of the speed of sound at freezing. At the same time, a unique value $\beta=1/3$ is sufficient to describe an extended dense region of the phase diagram, which makes the proposed approach particularly appealing.  


The chosen functional form of Eq.~(\ref{model}) is just a simple empirical approximation and is, of course, not unique. However, there are physical arguments behind the chosen form. In a two-phase motivated model~\cite{MoonPRR2024}, one can assume that a fluid can be approximated by a superposition of $x$ ideal gas atoms and $1-x$ liquidus atoms. The gas-like atoms are characterized by the sound velocity $c_{\rm s}\simeq \sqrt{\gamma}v_{\rm T}$, the liquidus atoms are characterized by the sound velocity $c_{\rm s}\simeq \alpha v_{\rm T}\sim 8v_{\rm T}$~\cite{RosenfeldJPCM_1999_HS,KhrapakPoF2023}. The reduced sound velocity of their superposition is then
\begin{equation}\label{2phase}
\frac{c_{\rm s}}{v_{\rm T}}\simeq x\sqrt{\gamma}+(1-x)\alpha.
\end{equation}
The main difficulty would be to determine the gas abundance $x$. In a recent application of the two-phase model to heat capacity, it has been proposed to relate this parameter to the ratio of unstable and stable instantaneous modes~\cite{MoonPRR2024}. The fluidity parameter $1-x$ has been shown to decay monotonically to zero as the temperature increases without apparent temperature scales involved. The most general scale-free function is a power-law function of the form $1-x\propto (T_{\rm fr}/T)^{\beta}$ for $T>T_{\rm fr}$~\cite{2phase}. Substituting this into Eq.~(\ref{2phase}) will lead us directly to Eq.~(\ref{model}) after minor redefinitions.       

The two dashed curves shown in Fig.~\ref{Fig1} correspond to the model of Eq.~(\ref{model}) with $\beta=1/3$ and $\alpha=6.8$ (lower curve) and $\alpha=7.2$ (upper curve). The agreement between the calculations and a simple empirical model is remarkable. The transport properties of some simple real fluids can be well reproduced using the LJ fluid model as a reference system. A recent example is the freezing density scaling of the transport coefficients applied to liquefied noble gases~\cite{KhrapakJCP2022_1} and methane~\cite{KhrapakJMolLiq2022}. It is therefore tempting to verify whether the freezing temperature scaling of the sound speed applies to real liquids as well. If this is the case, then the additional question is how are the parameters $\alpha$ and $\beta$ related to those of the LJ fluid.             

\section{Real liquids}

Let us check whether the scaling operating in the LJ fluid applies to real liquids. Naturally, we start with liquefied noble gases and then move to molecular liquids such as nitrogen and methane. Data are taken from the National Institute of Standards and Technology (NIST) Reference Fluid Thermodynamic and Transport Properties Database (REFPROP 10.0)~\cite{Refprop}. REFPROP calculates various thermodynamic and transport properties of industrially important fluids and their mixtures, based on models built on a foundation of experimental data~\cite{Refprop}. For details about the models for EoS and phase boundaries of various pure liquids implemented in REFPROP 10.0, see Refs.~\cite{Refprop,Huber2018}. Note that the models used are generally more advanced than those based on the LJ interaction potential with appropriate substance-dependent length and energy scales.

\subsection{Liquefied noble gases}

\begin{figure}
\includegraphics[width=8cm]{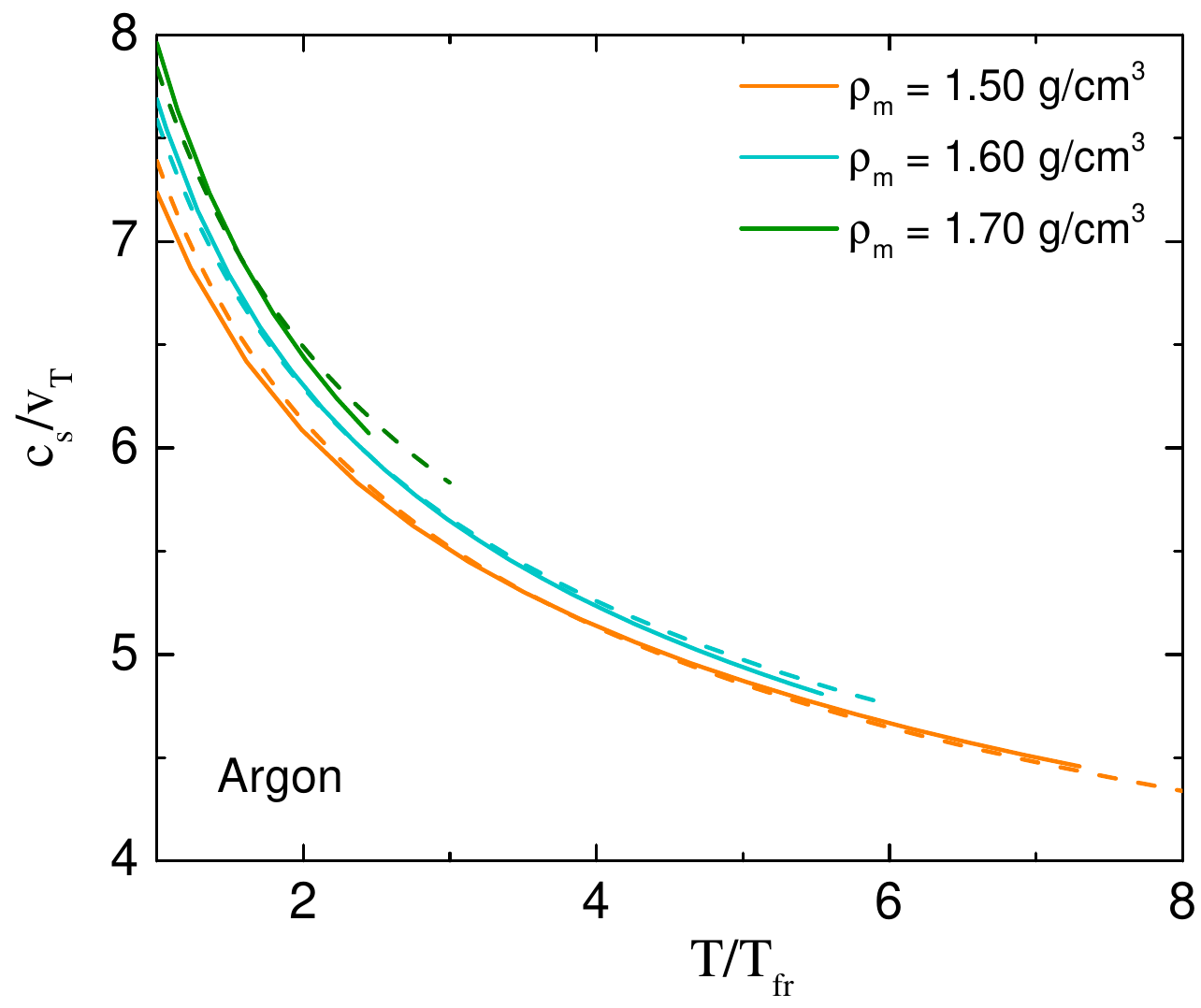}
\caption{(Color online) Reduced sound speed in liquefied argon along three isochores $\rho_m=1.5$, 1.6 and 1.7 g/cm$^3$ (from bottom to top). The solid curves mark the recommended values from REFPROP 10.0 database~\cite{Refprop}. The dashed curves are plotted using Eq.~(\ref{model})  with $\alpha=6.1$, 6.3, 6.55 (from bottom to top) and a unique $\beta=1/3$.  }
\label{Fig2}
\end{figure}

\begin{figure}
\includegraphics[width=8cm]{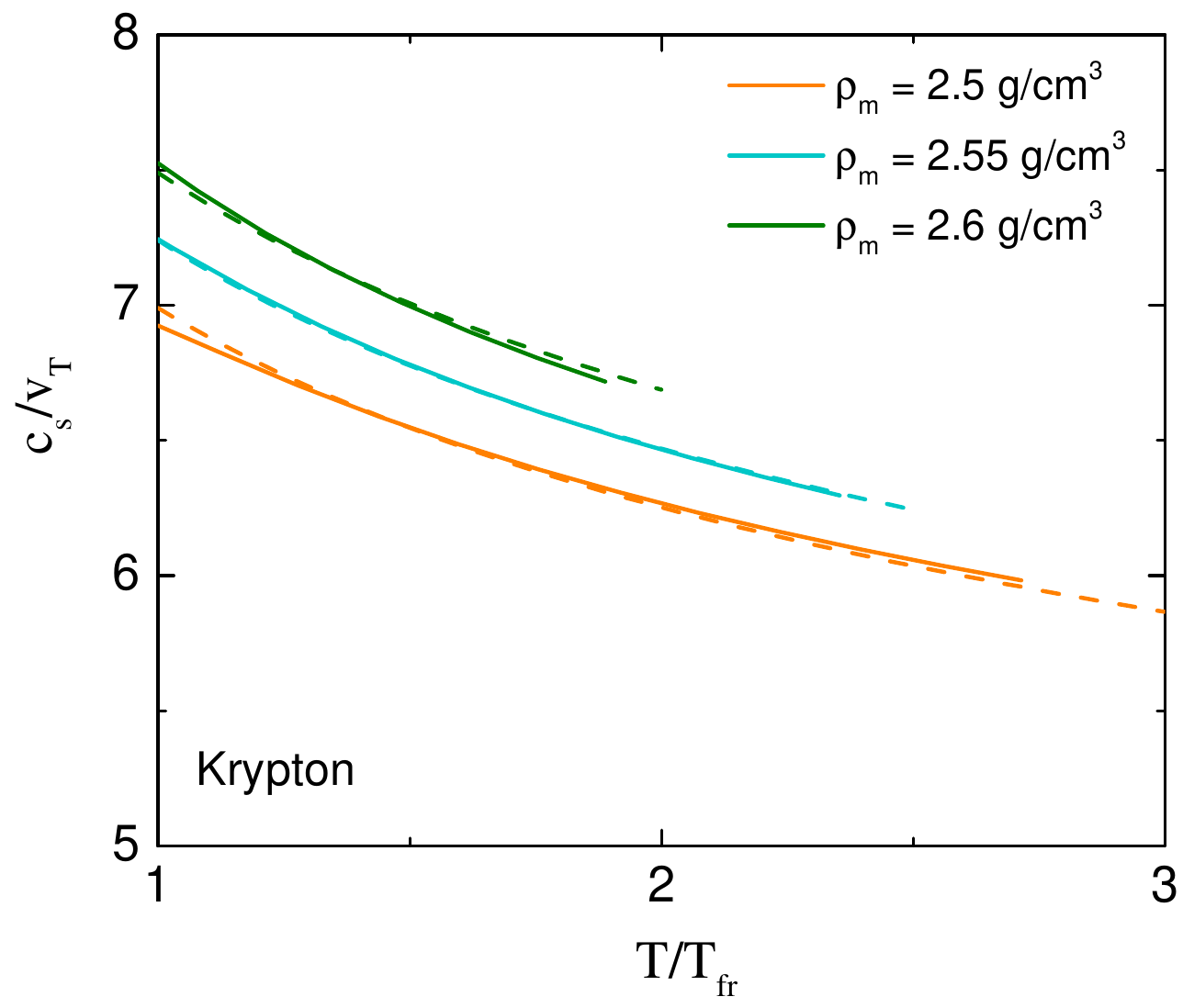}
\caption{(Color online) Reduced sound speed in liquefied krypton along three isochores $\rho_m=2.5$, 2.55 and 2.6 g/cm$^3$ (from bottom to top). The solid curves mark the recommended values from REFPROP 10.0 database~\cite{Refprop}. The dashed curves are plotted using Eq.~(\ref{model})  with $\alpha=5.7$, 5.95, 6.2 (from bottom to top) and a unique $\beta=1/5$. }
\label{Fig3}
\end{figure}

\begin{figure}
\includegraphics[width=8cm]{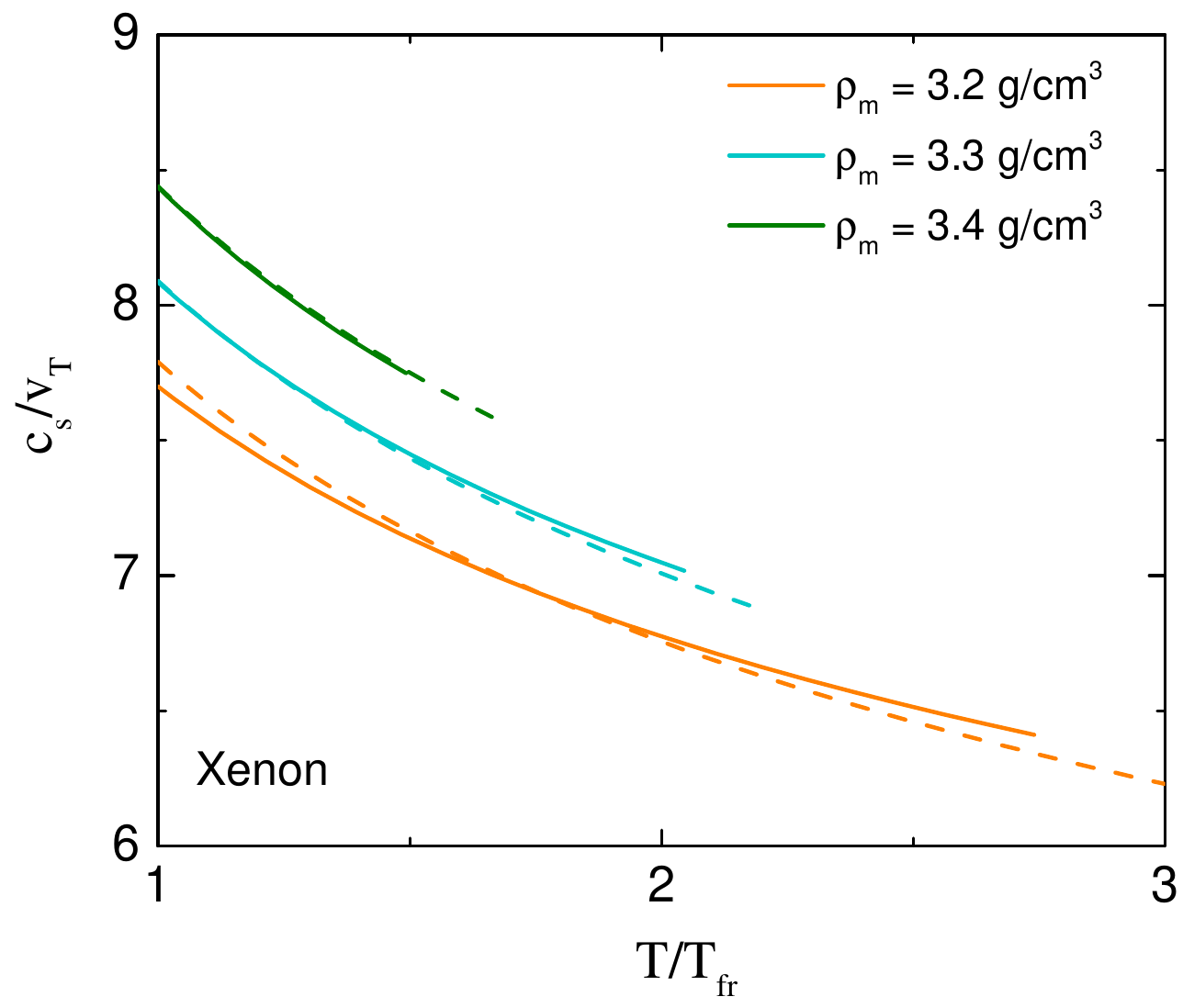}
\caption{(Color online) Reduced sound speed in liquefied xenon along three isochores $\rho_m=3.2$, 3.3 and 3.4 g/cm$^3$ (from bottom to top). The solid curves mark the recommended values from REFPROP 10.0 database~\cite{Refprop}. The dashed curves are plotted using Eq.~(\ref{model})  with $\alpha=6.5$, 6.8, 7.15 (from bottom to top) and a unique $\beta=1/4$.  }
\label{Fig4}
\end{figure}

Monoatomic liquefied noble gases argon (Ar), krypton (Kr), and xenon (Xe) are considered in this Section. Neon (Ne) is not considered because quantum effects can be important in this case~\cite{HansenPR1969,KhrapakPoF2023}. The results are shown in Figs.~\ref{Fig2} --\ref{Fig4}. The solid curves correspond to the recommended data from REFPROP 10.0. The dashed curves are plotted using Eq.~(\ref{model}). The heat capacity ratio for monatomic liquids is taken in all cases as $\gamma=5/3$, just as for the LJ fluid. The parameter $\alpha$ that governs the reduced sound velocity at the freezing transition increases with density, as expected from the LJ case. Typical values of reduced sound velocities near freezing are also comparable, $c_{\rm s}/v_{\rm T}\sim 7.5$ for argon, $c_{\rm s}/v_{\rm T}\sim 7$ for krypton, and $c_{\rm s}/v_{\rm T}\sim 8$ for xenon. The quasi-universality of reduced sound velocity at freezing is a general property of simple fluids with sufficiently steep interaction potentials~\cite{RosenfeldJPCM_1999_HS,KhrapakPoF2023}.

In the investigated parameter regime, a unique value of $\beta$ is sufficient to describe the dependence of the reduced speed of sound on temperature for each of the liquids considered. However, the exponent $\beta$ can vary considerably from one liquid to another. In argon, $\beta = 1/3$ as in the LJ fluid. In krypton and xenon, the reduced sound velocity drops slower with temperature with $\beta= 1/5$ in krypton and $\beta = 1/4$ in xenon, respectively. This implies that the exponent $\beta$ in Eq.~(\ref{model}) is not fully universal. Rather, it should be considered as a material-dependent property. This resembles the situation with the generalized Rosenfeld-Tarazons scaling of the heat capacity in simple fluids~\cite{KhrapakPOF11_2024}.          

\subsection{Molecular liquids}

\begin{figure}
\includegraphics[width=8cm]{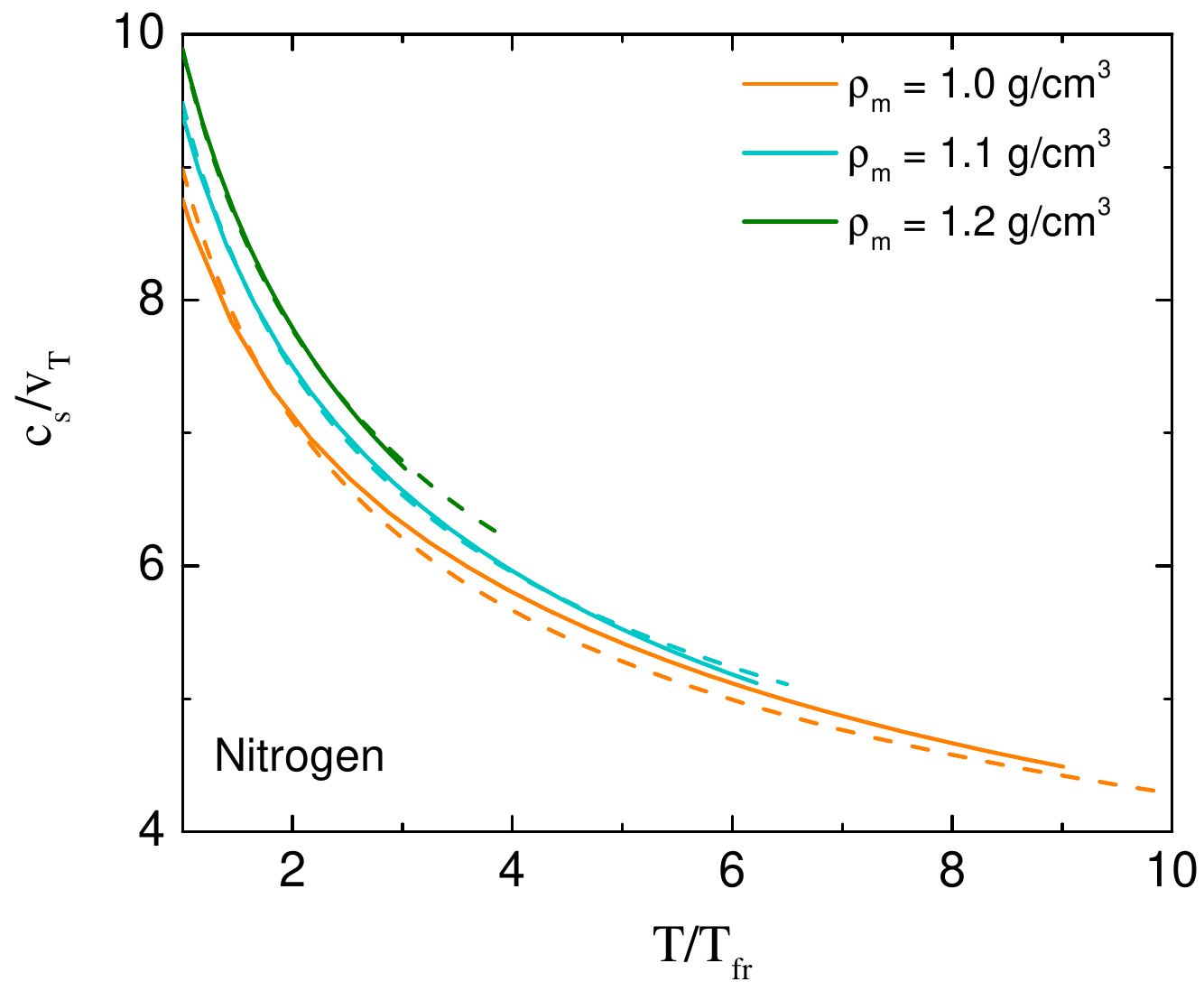}
\caption{(Color online) Reduced sound speed in liquefied nitrogen N$_2$ along three isochores $\rho_m=1$, 1.1, and 1.2 g/cm$^3$ (from bottom to top). The solid curves mark the recommended values from REFPROP 10.0 database~\cite{Refprop}. The dashed curves are plotted using Eq.~(\ref{model})  with $\alpha=7.8$, 8.3, and 8.7 (from bottom to top) and a unique value of  $\beta=2/5$. The fixed heat capacity ratio is $\gamma=1.4$.}
\label{Fig5}
\end{figure}

\begin{figure}
\includegraphics[width=8cm]{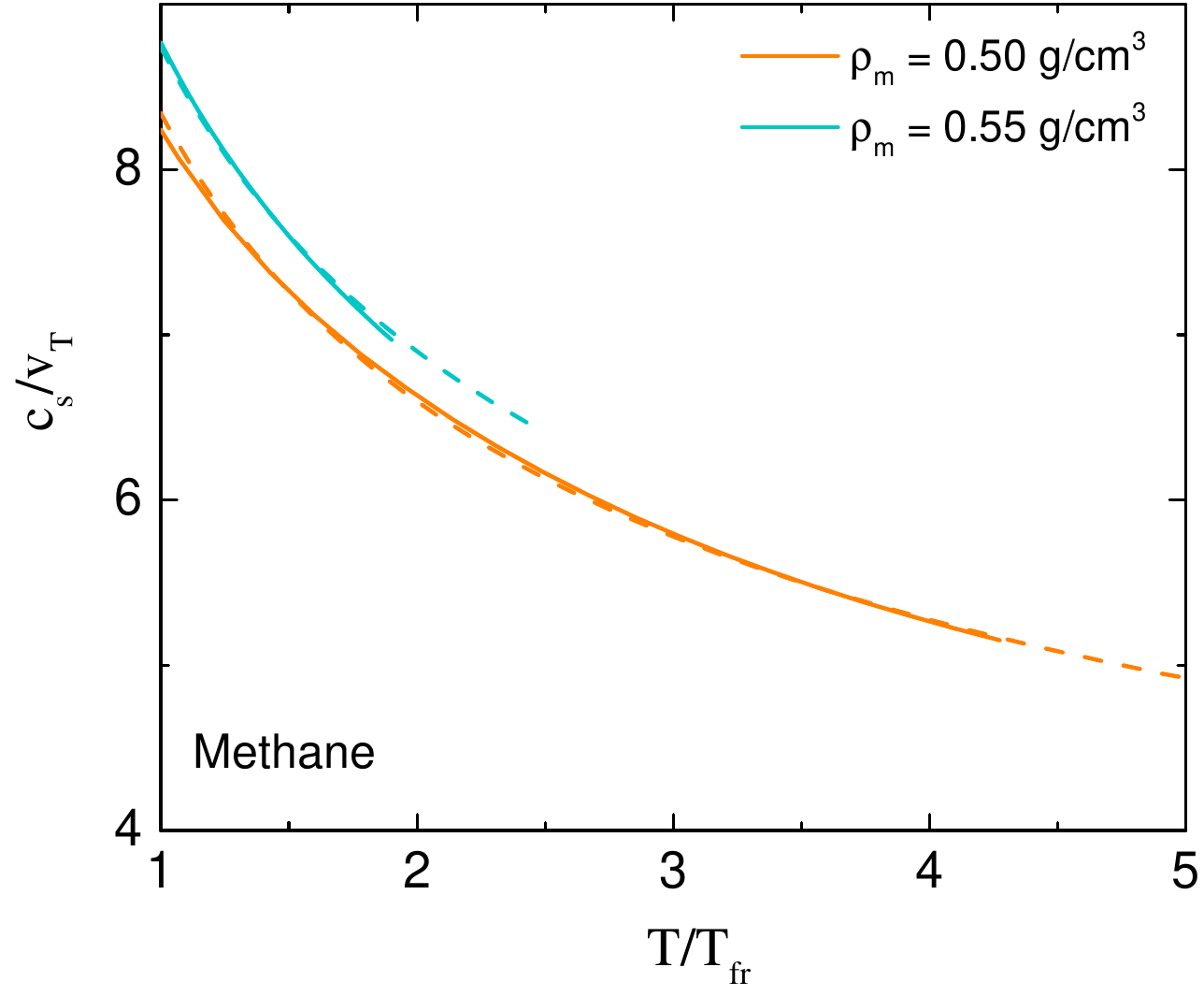}
\caption{(Color online) Reduced sound speed in liquefied methane along two isochores $\rho_m=0.5$ and 0.55 g/cm$^3$ (from bottom to top). The solid curves mark the recommended values from REFPROP 10.0 database~\cite{Refprop}. The dashed curves are plotted using Eq.~(\ref{model})  with $\alpha=7.2$ and 7.6 (from bottom to top) and a unique value of  $\beta=2/5$. The fixed heat capacity ratio is $\gamma=1.3$.}
\label{Fig6}
\end{figure}

Concerning molecular liquids, we consider nitrogen and methane. The results are summarized in Figs.~\ref{Fig5} and \ref{Fig6}. The heat capacity ratio for nitrogen gas is taken as $\gamma = 7/5=1.4$ corresponding to three translational degrees and two rotational degrees of freedom. For the ideal gas limit of methane, we take $\gamma = 1.3$. Some deviations from these values are possible, depending on the exact location on the phase diagram, but are not essential for the present consideration.   

We observe that the practical model of Eq.~(\ref{model}) describes well the recommended data from the REFPROP 10.0 database. For the molecular liquids considered, the reduced sound velocity at the freezing transition remains close to that in monatomic liquids $c_{\rm s}/v_{\rm T}\sim 8$. Hence, the values of the parameter $\alpha$ are also close. Finally, the same value of the exponent $\beta = 2/5$ allows us to describe the data for both nitrogen and methane liquids very well.  

For completeness and to simplify practical application of the obtained results we summarize the parameters used to fit the LJ and real fluids data in Tab.~\ref{Tab1} of the Appendix.

\section{Methane under extreme conditions}

In a recent experiment, the speed of sound in methane has been measured experimentally under conditions of the planetary interior~\cite{WhitePRR2024}. Specifically, the sound speed of warm dense methane created by laser heating a cryogen liquid jet has been measured at a temperature of 3480 K ($0.3$ eV) and a mass density of $\simeq 0.8$ g/cm$^3$. The derived sound speed is $5.9\pm 0.5$ km/s, which provides a high-temperature reference data point for methane. Under these conditions significant ionization is inevitable. However, the derived speed of sound is consistent with Birch's law, which is based on the data obtained at room temperature~\cite{WhitePRR2024}.    

Here we provide a speculative estimate, based on the freezing-temperature scaling of the sound speed in methane. The main steps of our approach are as follows. We consider a hypothetical supercritical methane molecular fluid at $T=3480$ K and $\rho_{m}=0.8$ g/cm$^3$ neglecting ionization and other high-temperature phenomena. First, we express the temperature and density at the state point considered in terms of the methane critical temperature and density. Then we locate the point with the same $T/T_{\rm c}$ and $\rho/\rho_{\rm c}$ on the LJ system phase diagram. Using an existing analytical approximation for the freezing curve of the LJ fluid in the form of $T_{\rm fr}^*=T_{\rm fr}^*(\rho^*)$ we can estimate the reduced temperature $T/T_{\rm fr}$ for the considered state point. Finally, using the freezing temperature scaling of the sound speed in methane, the actual sound speed of the idealized system is estimated.  

Each of these steps can introduce some inaccuracy in the final result. For example, the principle of corresponding states is known to work better for atomic substances than for molecular ones~\cite{GuggenheimJCP1945}. However, no better alternative appears to be available at this time. Our result can still provide some indications about the effect of ionization on the sound speed.    


The details of this calculation are as follows. The temperature and density at the critical point of methane are $T_c\simeq 190.55$ K and $\rho_c\simeq 0.163$ g/cm$^3$~\cite{FriendJPCRD1989}. The investigated state point with the density $0.8$ g/cm$^3$ thus corresponds to $\rho/\rho_c\simeq 4.9$ (note that the maximum density covered in the REFPROP 10.0 database corresponds to 0.6 g/cm$^3$). The critical point in the LJ fluid is located at $T^*_c\simeq 1.32$ and $\rho_c^*\simeq 0.31$ in conventional LJ units~\cite{Thol2016}. Translated into LJ fluid, the condition $\rho/\rho_c\simeq 4.9$ implies $\rho^*\simeq 1.5$. We can then use an approximate equation for the freezing curve in the LJ fluid~\cite{KhrapakJCP2011_2},
\begin{displaymath}
T_{\rm fr}^*\simeq 2.166 (\rho^*)^4-0.581(\rho^*)^2,
\end{displaymath}
to get $T^*_{\rm fr}\simeq 10.0$ at $\rho^*_{\rm fr}=1.5$. This corresponds to $T_{\rm fr}/T_c\simeq 7.6$ in the LJ fluid. Returning to the methane fluid, we obtain an estimate $T_{\rm fr}\simeq 7.6T_c\simeq 1450$ K. This means that the reduced temperature at the investigated state point is $T/T_{\rm fr}\simeq 3480/1450\simeq 2.4$. Applying now Eq.~(\ref{model}) with $\gamma = 1.3$, $\alpha = 8$, and $\beta=2/5$ we get $c_s/v_{\rm T}\simeq 6.8$. The thermal velocity of methane molecules at $T=3480$ K is $v_{\rm T}\simeq 1.3$ km/s. This gives our final result 
\begin{equation}\label{estimate}
c_{\rm s}\simeq 9.1 \quad {\rm km/s}.
\end{equation}
The experimentally determined speed of sound is $5.9\pm 0.5$ km/s, that is, about 30\% lower than our current estimate. This seems to indicate that ionization can significantly suppress the sound speed. However, it should be reminded that, in addition to neglecting ionization, our simple estimate can potentially be subject to inaccuracies related to the actual location of the freezing point at a density $0.8$ g/cm$^3$ in methane and the application of the freezing temperature scaling beyond the tested regime.    

\section{Conclusion}

The speed of sound is a very important property of materials and it is highly desirable to be able to predict it in different situations. Accurate prediction requires knowledge of an accurate equation of state. The latter is not always available or not in the entire domain of the phase diagram. To overcome this, in this paper we propose a freezing-temperature scaling of the reduced sound velocity in dense liquids.

The scaling is given by Eq.~(\ref{model}), which represents a linear superposition of the sound velocity in the ideal gas limit and the excess component exhibiting the power-law freezing temperature scaling. It is demonstrated that this scaling works well in the dense LJ fluid along isochores with densities above that at the triple point. It is then verified that similar scaling operates in atomic (argon, krypton, xenon) and molecular (nitrogen, methane) liquids. The fitting parameters are tabulated. This provides a simple practical way to estimate the speed of sound in dense fluids in parameter regimes where no experimental data is yet available.  

The freezing temperature scaling combined with the principle of corresponding states is applied to estimate the speed of sound in methane at an elevated temperature and density and to compare it with a recent experimental measurement. This comparison provides a preliminary indication that ionization can reduce the sound speed compared to the case of a non-ionized liquid at the same temperature and density. However, more data are needed to conclusively verify this trend.

Altogether, the obtained results can help to better understand the magnitude and main trends experienced by the sound speed of various atomic and molecular liquids at sufficiently high temperatures and densities. 

The authors declare no conflict of interests.

The data that support the findings of this study are available from the authors upon reasonable request. 

\appendix

\section{Fitting parameters}
The fitting parameters $\alpha$, $\beta$, and $\gamma$ used to fit various sound speed data with the help of Eq.~(\ref{model}) are summarized in Tab.~\ref{Tab1}. To simplify the comparison between different substances, the mass density at the isochore is translated into the ratio of the temperature on the liquidus to the triple point temperature. The corresponding reduced sound speeds at the liquidus are also provided. The dependence of the reduced sound speed on the ratio $T/T_{\rm tp}$ along the freezing curve of 15 atomic and molecular liquids is shown in Fig.~4 of Ref.~\cite{KhrapakPoF2023}.     
\begin{table}
\caption{\label{Tab1} Fitting parameters $\alpha$, $\beta$, and $\gamma$ in Eq.~(\ref{model}) used to fit the sound speed data shown in Figs.~\ref{Fig1} -- \ref{Fig6}. For the LJ fluid, the data corresponding to isochores $\rho^*=1.002$, 1.065, and 1.161 are included. Density is translated into the ratio of the temperature on the freezing curve to the triple point temperature, $T_{\rm tp}$. The actual reduced sound speed at the freezing curve is provided in the last column. }
\begin{ruledtabular}
\begin{tabular}{crrrrr}
System & $T/T_{\rm tp}$ & $\alpha$ & $\beta$ & $\gamma$ & $c_{\rm s}/v_{\rm T}$   \\ \hline
LJ & 2.16 & 6.8 & 1/3 & 5/3 & 8.09 \\
LJ  & 2.88 & 7.0 & 1/3 & 5/3  & 8.34\\
LJ  & 4.32 & 7.2 & 1/3 & 5/3  & 8.52\\
Argon & 1.26 & 6.1  & 1/3 & 5/3 & 7.24\\
Argon & 1.68 & 6.3  & 1/3 & 5/3 & 7.69\\
Argon & 2.19 & 6.55  & 1/3 & 5/3 & 7.96\\
Krypton & 1.08 & 5.7  & 1/5 & 5/3 & 6.92\\
Krypton & 1.17 & 5.95  & 1/5 & 5/3 & 7.24\\
Krypton & 1.28 & 6.2  & 1/5 & 5/3 & 7.53\\
Xenon & 1.40 & 6.5  & 1/4 & 5/3 & 7.70\\
Xenon & 1.61 & 6.8  & 1/4 & 5/3 & 8.08\\
Xenon & 1.87 & 7.15  & 1/4 & 5/3 & 8.44\\
Nitrogen & 1.76 & 7.8  & 2/5 & 1.4 & 8.76\\
Nitrogen & 2.49 & 8.3  & 2/5 & 1.4 & 9.39\\
Nitrogen & 3.35 & 8.7  & 2/5 & 1.4 & 9.89\\
Methane & 1.55 & 7.2  & 2/5 & 1.3 & 8.24\\
Methane & 2.20 & 7.6  & 2/5 & 1.3 & 8.76\\
\end{tabular}
\end{ruledtabular}
\end{table}



\bibliography{SE_Ref}

\end{document}